\documentclass{article}
\usepackage{spconf,amsmath,graphicx,hyperref}

\usepackage{color}
\usepackage{amsmath}
\usepackage{comment}
\usepackage{multirow}
\usepackage{amsfonts}
\usepackage{colortbl}
\usepackage{cite}
\usepackage{arydshln}
\usepackage{url}
\usepackage{mathrsfs}

\usepackage{amssymb}
\usepackage{amsfonts}
\usepackage{mathtools}
\usepackage{graphics}

\usepackage[table]{xcolor}
\newcommand{\red}[1]{{\textcolor{red}{#1}}}

\newcommand{\gr}[1]{{\textcolor[gray]{0.6}{#1}}}
\newcommand{\bm}[1]{{\mbox{\boldmath $#1$}}}
\newcommand{\ten}[1]{ \boldsymbol{\mathcal #1}}

\usepackage{graphicx}
\usepackage{subcaption} 

\title{Event classification by physics-informed inpainting for distributed multichannel acoustic sensor with partially degraded channels}
\name{\begin{tabular}{c}Noriyuki Tonami$^{1}$, Wataru Kohno$^{2}$, Yoshiyuki Yajima$^{1}$,\\ Sakiko Mishima$^{1}$, Yumi Arai$^{1}$, Reishi Kondo$^{1}$, Tomoyuki Hino$^{1}$\end{tabular}}
\address{$^{1}$NEC Corporation, Japan, $^{2}$NEC Laboratories America, Inc., The United States of America}
    
\begin{document}
\ninept

\maketitle

\begin{abstract}
Distributed multichannel acoustic sensing (DMAS) enables large-scale sound event classification (SEC), but performance drops when many channels are degraded and when sensor layouts at test time differ from training layouts. 
We propose a learning-free, physics-informed inpainting frontend based on reverse time migration (RTM). 
In this approach, observed multichannel spectrograms are first back-propagated on a 3D grid using an analytic Green’s function to form a scene-consistent image, and then forward-projected to reconstruct inpainted signals before log–mel feature extraction and transformer-based classification.
We evaluate the method on ESC-50 with 50 sensors and three layouts (circular, linear, right-angle), where per-channel SNRs are sampled from $-30$ to $0$ dB. 
Compared with an AST baseline, scaling-sparsemax channel selection, and channel-swap augmentation, the proposed RTM frontend achieves the best or competitive accuracy across all layouts, improving accuracy by $13.1$ points on the right-angle layout (from $9.7\%$ to $22.8\%$). 
Correlation analyses show that spatial weights align more strongly with SNR than with channel--source distance, and that higher SNR--weight correlation corresponds to higher SEC accuracy. 
These results demonstrate that a reconstruct-then-project, physics-based preprocessing effectively complements learning-only methods for DMAS under layout-open configurations and severe channel degradation.
\end{abstract}

\begin{keywords}
sound event classification, physics-informed, distributed acoustic sensing, reverse time migration
\end{keywords}

\section{Introduction}

Large-scale acoustic sensing and recognition technologies pave the way for next-generation environmental monitoring systems. 
Recent advances in sensing hardware, such as wireless acoustic sensor networks (WASN) \cite{Ntalampiras_ETCI2018,Ruiz_TASLP2022}, ad-hoc microphone arrays \cite{Gaubitch_TASLP2013,chen_interpseech2021}, and distributed fiber-optic sensors (DFOS) \cite{Lu_ICASSP2021,Ip_JOCN2022}, have enabled the acquisition of large-scale acoustic data from diverse locations, forming distributed multichannel acoustic sensing (DMAS). 
Leveraging DMAS, sound event classification (SEC) \cite{Gemmeke_ICASSP2017,Mesaros_SPmaga2021_01} can be applied to ambient signals to automatically detect and categorize acoustic events.
This leads to large-scale environmental monitoring systems capable of comprehensive ecosystem observation, real-time public safety surveillance, and sustainable urban soundscape management.

However, deploying DMAS with SEC in real environments faces three major challenges.
First, sensing channels are often partially degraded due to local conditions such as noise, hardware variability, or clipping \cite{Ip_JOCN2022}. 
Second, sensor layouts at test time may differ from those used during training. 
Machine-learning-based models inherently face the out-of-distribution (OOD) problem; hence, SEC models trained on one layout often degrade on others, requiring costly retraining. 
Third, the spatial scale between sensors and sources is often several meters or more, introducing synchronization issues, propagation delays, spatial aliasing, and reduced spatial coherence. 
Although sound event localization and detection (SELD) \cite{Politis_TASLP2020} has been studied, most work addresses only proximate sources within about 10 m. 
Overall, the key difficulty is that multichannel observations remain bound to sensor coordinates, limiting generalization across degraded channels and unseen layouts.

For the degraded channels and unseen layouts, a solution is to transform sensor-dependent observations into a geometry-independent physical representation without machine-learning-based methods. 
Within DMAS, DFOS has been studied with physics-based methods in seismology, where Reverse Time Migration (RTM) \cite{Hua_Wei_ESR2018,Kang_sensors2023} is a well-established imaging technique. 
RTM back-propagates observed signals through an analytic Green’s function to construct a scene-consistent image on a common grid, removing direct dependence on sensor coordinates. 
Forward projection from this image then reconstructs inpainted signals with equalized quality across channels.

To address the realistic problem of partially degraded channels and layout-open in DMAS-based sound event classification, we propose a physics-informed-inpainting frontend utilizing RTM.
In the proposed method, we reconstruct multichannel signals before classifying events.
The proposed method fills in missing or low SNR channels without machine learning, with constraints of physics.

\vspace{0pt}
\section{Preliminaries}
\vspace{0pt}

\subsection{Reverse time migration for imaging}


We consider the observed multichannel spectrogram $\mathbf{X}\in\mathbb{C}^{N\times F\times T}$ with channels $n=1 \dots N$, frequency bins $f=1 \dots F$, and frames $t=1 \dots T$.
The purpose of reverse time migration is to obtain an image map $\mathbf{M}\in\mathbb{C}^{J\times F\times T}$ defined on grid points $\bm{g}_j\in\mathbb{R}^3$ and sensor positions $\bm{s}_n\in\mathbb{R}^3$.
\begin{align}
r_{n j} &= \lVert \bm{s}_n - \bm{g}_j \rVert_2 \in \mathbb{R}^{+},
&
k_f &= \frac{\omega_f}{c} \in \mathbb{R}^{+}.
\label{eq:geom}
\end{align}
\noindent
Here $r_{n j}$ denotes the sensor–grid distance and $k_f$ the wavenumber at frequency bin $f$.
We define the propagation operator $\mathbf{L}\in\mathbb{C}^{F\times N\times J}$ by the analytic three-dimensional free-space Green’s function, which carries spherical-wave amplitude and phase.
\begin{align}
L_{f n j}
&= \frac{\exp\!\bigl(i\,k_f\, r_{n j}\bigr)}{4\pi\, r_{n j}} \in \mathbb{C}.
\label{eq:L-3d}
\end{align}

We subsequently obtain $\mathbf{M}$ using $\mathbf{L}$ and $\mathbf{X}$.
For each frequency $f$ and frame $t$, image components $M_{j f t}$ on the grid are obtained by back-propagating $\mathbf{X}$ with the conjugated kernel.
\begin{align}
M_{j f t}
&= \sum_{n=1}^{N} \overline{L_{f n j}}\, X_{n f t},
\label{eq:rtm}
\end{align}

\begin{figure}[t!]
  \centering
  \includegraphics[width=0.5\textwidth]{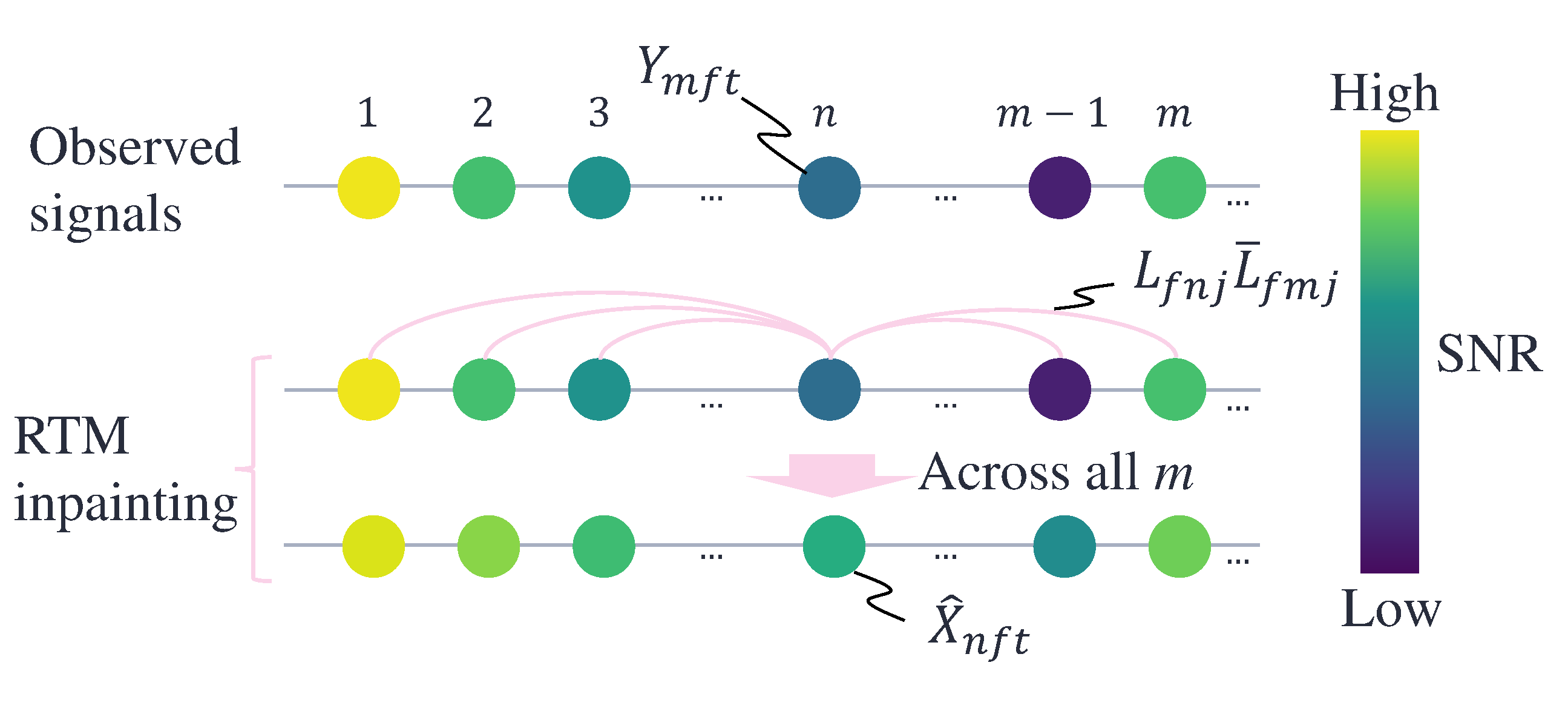}\\
  \vspace{-10pt}
  \caption{Concept of multichannel signals with partially degraded channels and RTM inpainting}
  \vspace{-0pt}
  \label{fig:ICASSP2026_prop_cencept_figure}
\end{figure}

\begin{figure}[t!]
  \centering
  \includegraphics[width=0.5\textwidth]{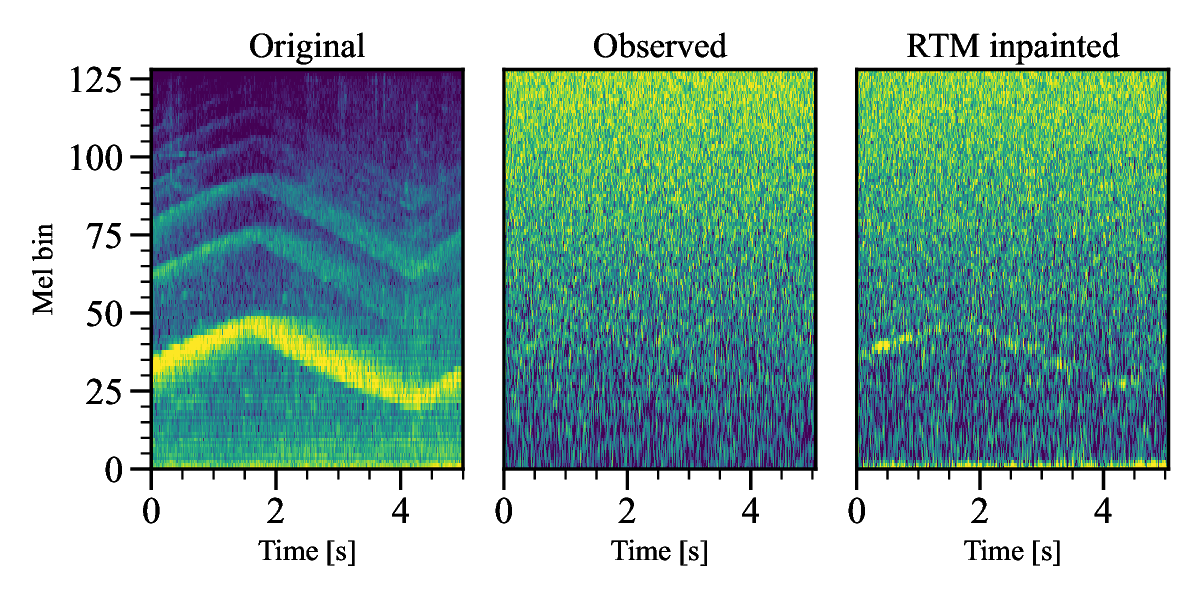}\\
  \vspace{-10pt}
  \caption{Example of RTM-inpaint spectrogram}
  \vspace{0pt}
  \label{fig:example_RTM_inpaint}
\end{figure}

\subsection{Multi-input single-output SEC}
With multichannel signals, multi-input single-output SEC is formalized as follows:
\begin{align}
\hat{e} = \underset{e}{\operatorname{argmax}} f(\ten{X})
\label{eq:MISO_SEC}
\end{align}
\noindent
where $\hat{e}$, $f$, and $\ten{X}$ denote an estimated sound event class, SEC model, and multichannel acoustic signals.
Most studies assume $\ten{X}$ has no partially degraded channels, which is not realistic.

\vspace{0pt}
\section{Proposed method}
\vspace{0pt}
As shown in Figure \ref{fig:ICASSP2026_prop_cencept_figure}, we first formulate our task, that is, DMAS-based SEC with partially degraded channels.
We then propose a learning-free and physics-informed frontend method to address the realistic problems of the partially degraded channels and layout-open.
We further formulate learnable spatial weights to comprehend the contributions of channels. 

\subsection{Multichannel SEC with degraded channels}
We first consider a degraded multichannel spectrogram $\mathbf{Y}\in\mathbb{C}^{N\times F\times T}$ and an unknown clean spectrogram $\mathbf{X}^{\star}\in\mathbb{C}^{N\times F\times T}$, with channel index $n=1 \dots N$, frequency bin $f=1 \dots F$, and frame $t=1 \dots T$.
Let $\mathcal{D}\subset\{1\dots N\}$ denote the degraded-channel set and $\mathcal{R}=\{1\dots N\}\setminus\mathcal{D}$ the reliable set.
Channel $n$ is declared degraded by $\mu_n$ with arbitrary threshold $\tau$, and the observation model distinguishes high- and low-noise components as follows:
\begin{align}
&n\in\mathcal{D}\ \Longleftrightarrow\ \mu_{n}<\tau, \\
Y_{nft}=&
\begin{cases}
X^{\star}_{nft}+\eta_{nft}, & n\in\mathcal{D},\\[4pt]
X^{\star}_{nft}+\nu_{nft},  & n\in\mathcal{R},
\end{cases}
\qquad
\mathbb{E}\!\bigl[|\eta_{nft}|^{2}\bigr]\gg \mathbb{E}\!\bigl[|\nu_{nft}|^{2}\bigr]. 
\end{align}
Here, $\eta_{nft}$ and $\nu_{nft}$ represent complex-value noise.
Using $\mathbf{Y}$ as an input of Eq. \ref{eq:MISO_SEC}, we operate the task of classifying sound events with partially degraded channels.

\begin{figure}[t!]
  \centering
  \subfloat[Coordinate of DMAS, which consists of 50 channels]{%
    \includegraphics[width=.99\linewidth]{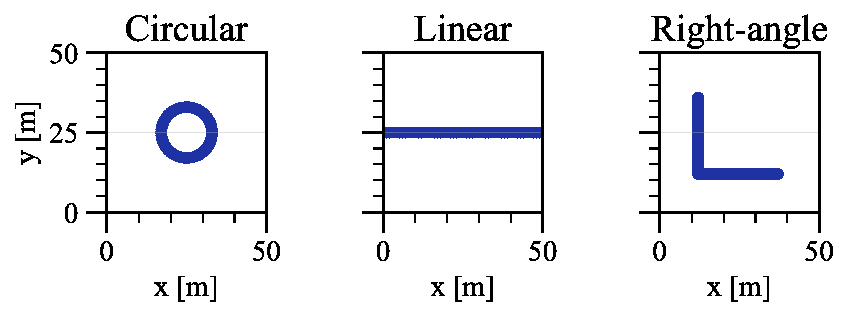}\label{fig:Coordinate}}

  \subfloat[Distribution of channel--source distances]{%
    \includegraphics[width=.99\linewidth]{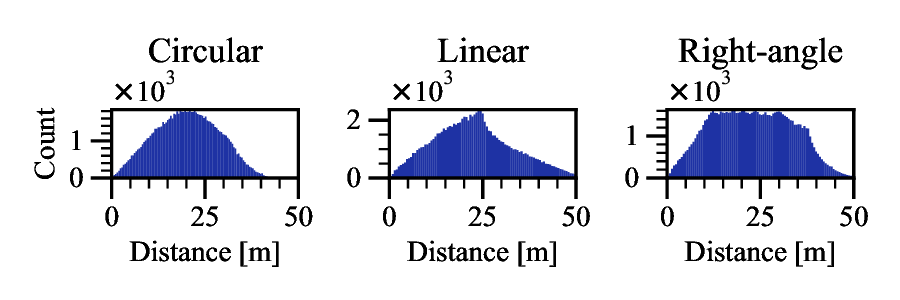}\label{fig:distribution_distance}}
  
  \caption{(a) Coordinate and (b) distance distribution}
  \label{fig:coord_dist_stack}
\end{figure}

\begin{figure*}[t!]
  \centering
  \includegraphics[width=1.0\textwidth]{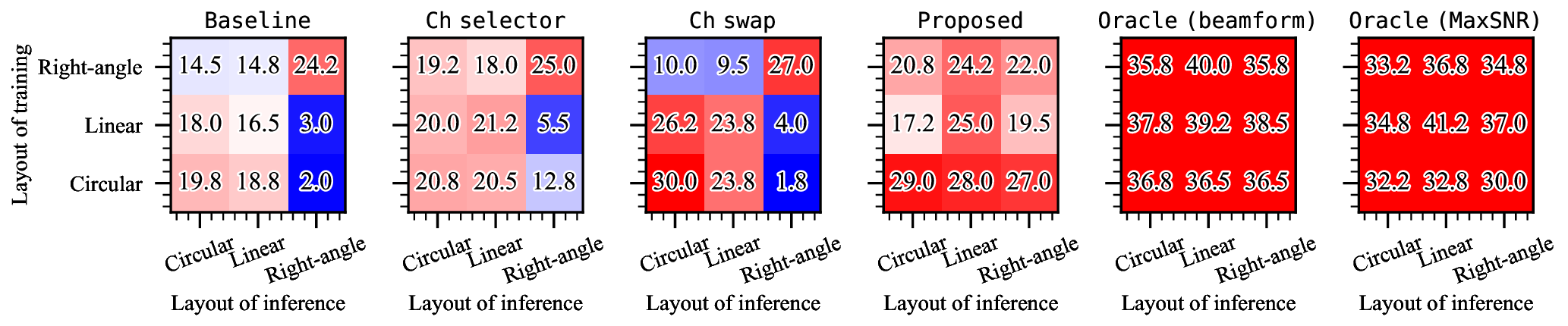}
  \vspace{-18pt}
  \caption{Accuracy [\%] of SEC for each layout}
  \vspace{-12pt}
  \label{fig:accuracy_layout}
\end{figure*}

\subsection{RTM-based inpainting}
Assume an RTM image $\mathbf{M}\in\mathbb{C}^{J\times F\times T}$ has been computed using all channels of $\mathbf{Y}$ with a propagation operator $\mathbf{L}\in\mathbb{C}^{F\times N\times J}$.
The inpainted spectrogram $\widehat{\mathbf{X}}\in\mathbb{C}^{N\times F\times T}$ is obtained by forward projection:
\begin{align}
\widehat{X}_{nft}&=\sum_{j=1}^{J} L_{fnj}\, M_{jft}\\
&= \sum_{m=1}^{N} \left( \sum_{j=1}^{J} L_{fnj}\,\overline{L}_{fmj} \right) Y_{mft}.
\label{eq:forward-inpaint}
\end{align}
\noindent
The operator takes the form $L_{fnj}\,\overline{L}_{fmj}$, which corresponds to a physics-informed spatial filter that theoretically accounts for wave propagation (e.g., speed of sound, sensor layout, distance between sensing channels, and wave attenuation). 
This enables the production of more equalized-quality multichannel data from the originally degraded observations, and in turn, leads to improved multi-channel SEC accuracy and greater robustness with respect to sensor layouts. 
In this paper, we consider the operator with Eq. \ref{eq:L-3d}, based on free-space Green's function for the Helmholtz-type wavefield equation. Meanwhile, it is also worth noting that other types of filters and/or simulations may be obtained by solving different wave scattering problems, e.g., solutions for multi-layered media \cite{ozdemir_JASA2022}. 
 
Figure \ref{fig:example_RTM_inpaint} shows an example of RTM-based inpainting, which is not just denoising.
$\widehat{X}_{n f t}$ is then projected onto log-mel spectrogram with a mel filterbank $\mathbf{A}\in\mathbb{R}^{K\times F}$:
\begin{align}
E_{n k t} &= \log_{10} \Biggl( \sum_{f=1}^{F} A_{k f}\, \bigl\lvert \widehat{X}_{n f t}  \bigr\rvert^{2}\Biggr).
\label{eq:mel-energy}
\end{align}

\subsection{Per-channel patch embedding to investigate contributions of channels}

For each channel \(n \in \{1,\dots,N\}\), we consider a sequence of \(I\) local time--frequency patches obtained by a fixed stride over the \(K \times T\) plane.
Let \(p_{n,i} \in \mathbb{R}^{K_p T_p}\) be the vectorized content of the \(i\)-th local patch on channel \(n\).
Each patch is projected to a \(D\)-dimensional token with a learned linear map
\(W_{\mathrm{emb}} \in \mathbb{R}^{(K_p T_p)\times D}\) and bias \(b_{\mathrm{emb}} \in \mathbb{R}^{D}\):
\begin{align}
t_{n,i}
&= p_{n,i}\, W_{\mathrm{emb}} + b_{\mathrm{emb}}
\in \mathbb{R}^{D},
~
(n=1,\dots,N;\; i=1,\dots,I).
\end{align}
\noindent
We then compress the \(N\) channel-wise tokens at each patch index by simple averaging:
\begin{align}
\bar{t}_{i}
&= \frac{1}{N}\sum_{n=1}^{N} t_{n,i}
\in \mathbb{R}^{D},
\qquad
(i=1,\dots,I).
\label{eq:per-channel patch embedding}
\end{align}
Collecting \(\bar{t}_{i}\) over \(i\) yields a single sequence
\(
\bar{T} \in \mathbb{R}^{I \times D}
\),
which is passed to single-channel SEC models, such as Audio Spectrogram Transformer \cite{gong_interspeech_2021}.

Moreover, we obtain spatial weights to investigate how the balance among channels impacts the SEC performance:
For each patch, we compute similarity scores $\langle \cdot \rangle$ by scaled inner products between \(\bar{t}_i\) and \(t_{n,i}\), and apply a softmax operation across the channel dimension.
Finally, we average these weights across patches to define the overall channel contribution:
\begin{align}
w_n
\;=\;
\frac{1}{I}\sum_{i=1}^{I}
\frac{\exp\!\left(\langle \bar{t}_i,\; t_{n,i}\rangle/\sqrt{D}\right)}{\sum_{m=1}^{N}\exp\!\left(\langle \bar{t}_i,\; t_{m,i}\rangle/\sqrt{D}\right)},
\qquad
\sum_{n=1}^{N} w_n = 1.
\label{eq:spatial_weight}
\end{align}
\noindent
The resulting vector \(\mathbf{w}=(w_1,\dots,w_N)\) can be used to investigate the relative contribution of each channel.

\begin{table}[t!]
\centering
\renewcommand{\arraystretch}{1.25}
\caption{Accuracy [\%] for each inference layout averaged over training layouts}
\label{tbl:overall_acc}
\vspace{-5pt}
\setlength{\tabcolsep}{6pt}
\scalebox{0.85}{
\begin{tabular}{lcccc}
\hline
\textbf{Method} & \textbf{Circular} & \textbf{Linear} & \textbf{Right-angle} & \textbf{Average} \\
\hline
{\tt Baseline}           & 17.4 & 16.7 &  9.7 & 14.6 \\
{\tt Ch selector}        & 20.0 & 19.9 & 14.4 & 18.1 \\
{\tt Ch swap}            & 22.1 & 19.0 & 10.9 & 17.3 \\
{\tt Proposed}           & \bf 22.3 & \bf 25.8 & \bf 22.8 & \bf 23.6 \\
\hdashline
\gr{\tt Oracle (beamform)} & \gr{36.8} & \gr{38.6} & \gr{36.9} & \gr{37.4} \\
\gr{\tt Oracle (MaxSNR)}   & \gr{33.4} & \gr{36.9} & \gr{33.9} & \gr{34.7} \\
\hline
\end{tabular}
}
\vspace{-8pt}
\end{table}

\begin{figure*}[t!]
  \centering
  \includegraphics[width=0.98\textwidth]{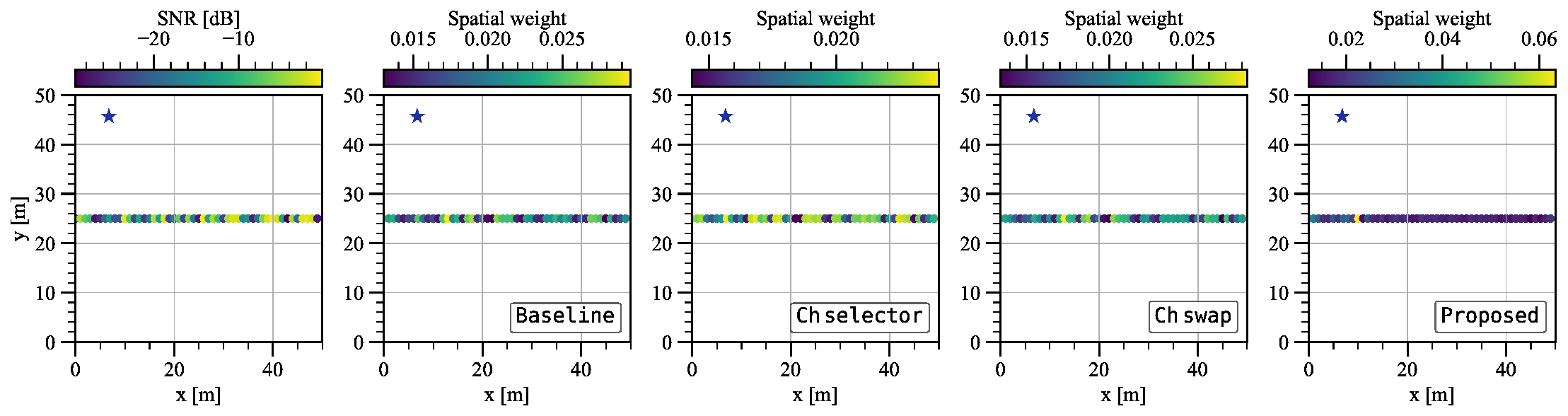}\\
  \vspace{-8pt}
  \caption{Visualization of spatial weight (Eq. \ref{eq:spatial_weight}) for each method}
  \vspace{-12pt}
  \label{fig:SNR_and_weights}
\end{figure*}

\begin{figure}[t!]
  \centering
  \includegraphics[width=0.45\textwidth]{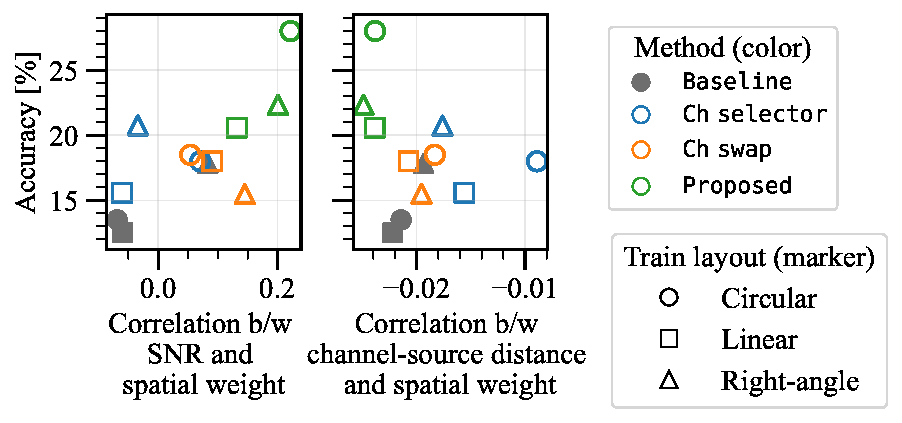}\\
  \vspace{-6pt}
  \caption{Relationships among accuracy of SEC, spatial weight (Eq. \ref{eq:spatial_weight}), SNR, and channel--source distance}
  \vspace{-10pt}
  \label{fig:scatter_acc_vs_corr}
\end{figure}

\vspace{0pt}
\section{Experiments}
\vspace{0pt}
\subsection{Experimental conditions}
\noindent
{\bf Dataset}:
To verify the performance of our method, we simulated large-scale distributed acoustic sensors with partially degraded channels.
In the simulation, we generated omnidirectional fifty-channel signals using the ESC-50 dataset \cite{piczak_ACMM2015}, where no reverberation and the speed of sound was set to 343 m/s.
ESC-50 comprises 2,000 audio signals with 50 environmental sound classes.
As shown in Figure \ref{fig:Coordinate}, we simulated three types of layouts of distributed acoustic sensors: circular, linear, and right-angle.
The spatial sampling of channels was set to one meter.
The location of a sound source $(x,y)$ was sampled from a uniform distribution: $x, y \sim [0, 50]$ meters, respectively.
The distribution of the distances between the sound sources and each channel is shown in Figure \ref{fig:distribution_distance}.
White Gaussian noises with variable SNRs were then added to the multichannel signals.
To reproduce partially degraded channels, the variable SNRs of each channel were sampled from $[-30, 0]$ dB.
Finally, we obtained 300,000 audio clips, where 2,000 sources $\times$ 50 channels $\times$ 3 layouts.
Following ESC-50, we conducted five-fold cross-validation.

\noindent
{\bf Comparison}:
\begin{itemize}
    \item {\tt Baseline}: Audio Spectrogram Transformer (AST) \cite{gong_interspeech_2021} with Eq. \ref{eq:per-channel patch embedding}.
    AST-based models have been widely used for SEC  \cite{koutini_interpseech2022,Elizalde_icassp2022,Chen_icml2023}.
    \item {\tt Ch selector}: {\tt Baseline} with scaling sparsemax based channel selection  \cite{chen_interpseech2021}. 
    \item {\tt Ch swap}: {\tt Baseline} with data augmentation of channel swapping \cite{imoto_EUSIPCO2021}.
    \item {\tt Proposed}: {\tt Baseline} using RTM-based inpainting with Eq. \ref{eq:forward-inpaint}.
    \\[5pt]
    Moreover, we employ oracle methods, which are expected to achieve the performance limit, as follows:
    \item {\tt Oracle (beamform)}: single-channel AST \cite{gong_interspeech_2021} with delay-and-sum beamforming using the ground truth of the sound source location.
    \item {\tt Oracle (MaxSNR)}: single-channel AST \cite{gong_interspeech_2021}  using the maximum-SNR channel obtained from the simulation configuration.
\end{itemize}
\noindent
For {\tt ch selector}, we implemented the selector before Eq. \ref{eq:per-channel patch embedding}.
For {\tt proposed}, $c$ is set to 343 m/s in Eq. \ref{eq:geom} with $\bf M$ on the x-y grid points $\bm{g}_j$ of 1$\times$1 m.
AST architecture is the same as VisionTransformer (ViT) Base \cite{Dosovitskiy_iclr2021}.
The models were optimized using Adam \cite{Kingma_ICLR2015} with an initial learning rate of 0.001 for 20 epochs.
In this study, all channels are assumed to be synchronized to focus on the problem of the partially degraded channels and the layout-open configuration.

\subsection{Experimental results}

\noindent
{\bf Overall accuracy across layouts}:\;
In this experiment, we confirm the accuracy of classifying events for each method as a summary result.
Table \ref{tbl:overall_acc} shows the SEC performances in terms of accuracy for each layout, where the scores are averaged over training layouts.
As shown in the table, the proposed SEC with RTM-based inpainting outperforms the conventional methods in terms of accuracy.
Of the inference layouts, the accuracy of the proposed method is significantly higher than that of the conventional methods for right-angle layout.
The proposed method improved the accuracy by 13.1 ($=22.8-9.7$) percentage points compared with {\tt baseline}.
This might be because the proposed RTM inpainting incorporates physics information that is robust to layout changes, unlike the conventional learning-based methods.
A remaining challenge is the significant performance gap between the proposed method and the oracles.

\noindent
{\bf Accuracy of SEC for each training and inference layout}:\;
Figure \ref{fig:accuracy_layout} shows the accuracy of SEC for each training and inference layout.
The result indicates that the proposed method maintains stable performance against layout changes compared with conventional methods.
{\tt Baseline}, {\tt ch selector}, and {\tt ch swap} resulted in significant deteriorations in terms of the accuracy of training and/or testing right-angle.
This is because those conventional methods overfit the linear and circular layouts that are different from the right-angle layout, as shown in Figure \ref{fig:distribution_distance}.
In particular, {\tt ch swap} tends to overfit in each group of \{circular, linear\} and right-angle.
While {\tt ch swap} encourages the model to learn geometry-invariant features, the right-angle layout, where channel--source distances for each channel are large on average, as shown in Figure \ref{fig:distribution_distance}, requires actively exploiting information on channels with small distances. 
Hence, this case, such as the right-angle layout, is less about layout generalization and more about layout adaptation.
The proposed method addresses this problem by improving SNR.

\noindent
{\bf Relationships among accuracy of SEC, spatial weight (Eq. \ref{eq:spatial_weight}), SNR, and channel--source distance}:\;
To further investigate how each SEC model employs information on the channel, we utilize the spatial weights formulated by Eq. \ref{eq:spatial_weight}.
Figure \ref{fig:scatter_acc_vs_corr} shows the relationships among the accuracy of SEC, the spatial weight (Eq. \ref{eq:spatial_weight}), SNR of the simulation, and the channel--source distance.
In the figure, we calculated Pearson's correlation coefficient as the correlation between the spatial weights and SNR or the channel--source distance.
Comparing the horizontal axis on left- and right-side scatter plots, we find that this relationship to SNR is stronger than that to the channel–source distance for the spatial weights.
Moreover, on the left side, the higher correlation with SNR is associated with higher accuracy.
The results indicate that the proposed method effectively exploits SNR-based information about which channels are most beneficial for SEC.

\noindent
{\bf Visualization of spatial weight (Eq. \ref{eq:spatial_weight}) for each method}:\;
Figure \ref{fig:SNR_and_weights} depicts an example of visualized spatial weights (Eq. \ref{eq:spatial_weight}).
The star mark represents the source location.
The left-side plot is for the simulated SNRs.
The others are for the visualizations of the spatial weight for each method.
This example shows the results obtained from SEC models trained by circular and tested by linear layouts, respectively.
The result clearly shows that the proposed method focuses on an informative channel in terms of SNR compared with the conventional methods.
With the results of {\tt oracle (MaxSNR)} from Table \ref{tbl:overall_acc} and Figure \ref{fig:accuracy_layout}, we find that selecting channels based on SNR criteria is crucial for the task of SEC with partially degraded channels.

\vspace{0pt}
\section{Conclusion}
\vspace{0pt}

We proposed a learning-free, physics-informed inpainting frontend based on RTM for SEC with distributed multichannel sensors under degraded channels and layout-open configurations. 
The method achieved the best accuracy across circular, linear, and right-angle layouts, with especially large gains in the right-angle case.
In particular, the proposed method improved classification accuracy by 13.1 percentage points compared with the conventional method.
The results showed that spatial weights align more strongly with SNR than with channel--source distance, and that higher SNR--weight correlation accompanies higher accuracy, indicating effective use of channel-quality cues while suppressing spurious evidence. 
In future work, we will investigate the SEC performance on asynchronous channels configuration.

\clearpage
\bibliographystyle{IEEEbib}
\bibliography{IEEEabrv,refs_et_al}

\begin{thebibliography}{10}

\bibitem{Ntalampiras_ETCI2018}
S.~Ntalampiras,
\newblock ``Moving vehicle classification using wireless acoustic sensor networks,''
\newblock {\em IEEE Transactions on Emerging Topics in Computational Intelligence}, vol. 2, no. 2, pp. 129--138, 2018.

\bibitem{Ruiz_TASLP2022}
S.~Ruiz, T.~Waterschoot, and M.~Moonen,
\newblock ``Distributed combined acoustic echo cancellation and noise reduction in wireless acoustic sensor and actuator networks,''
\newblock {\em IEEE/ACM Transactions on Audio, Speech, and Language Processing {\rm (}TASLP{\rm )}}, vol. 30, pp. 534--547, 2022.

\bibitem{Gaubitch_TASLP2013}
N.~Gaubitch, B.~Kleijn, and R.~Heusdens,
\newblock ``Auto-localization in ad-hoc microphone arrays,''
\newblock {\em Proc. {IEEE} International Conference on Acoustics, Speech and Signal Processing {\rm (}ICASSP{\rm )}}, pp. 106--110, 2013.

\bibitem{chen_interpseech2021}
J.~Chen and X.~Zhang,
\newblock ``Scaling sparsemax based channel selection for speech recognition with ad-hoc microphone arrays,''
\newblock {\em Proc. INTERSPEECH}, pp. 291--295, 2021.

\bibitem{Lu_ICASSP2021}
Y.~Lu, Y.~Tian, S.~Han, E.~Cosatto, S.~Ozharar, and Y.~Ding,
\newblock ``Automatic fine-grained localization of utility pole landmarks on distributed acoustic sensing traces based on bilinear resnets,''
\newblock {\em Proc. {IEEE} International Conference on Acoustics, Speech and Signal Processing {\rm (}ICASSP{\rm )}}, pp. 4675--4679, 2021.

\bibitem{Ip_JOCN2022}
E.~Ip, J.~Fang, Y.~Li, Q.~Wang, M.~Huang, M.~Salemi, and Y.~Huang,
\newblock ``Distributed fiber sensor network using telecom cables as sensing media: technology advancements and applications,''
\newblock {\em Journal of Optical Communications and Networking {\rm (}JOCN{\rm )}}, vol. 14, no. 1, pp. 61--68, 2022.

\bibitem{Gemmeke_ICASSP2017}
J.~Gemmeke, D.~Ellis, D.~Freedman, A.~Jansen, W.~Lawrence, R.~Moore, M.~Plakal, and M.~Ritter,
\newblock ``Audio set: An ontology and human-labeled dataset for audio events,''
\newblock {\em Proc. {IEEE} International Conference on Acoustics, Speech and Signal Processing {\rm (}ICASSP{\rm )}}, pp. 776--780, 2017.

\bibitem{Mesaros_SPmaga2021_01}
A.~Mesaros, T.~Heittola, T.~Virtanen, and M.~D. Plumbley,
\newblock ``Sound event detection: A tutorial,''
\newblock {\em {IEEE} Signal Processing Magazine}, vol. 38, no. 5, pp. 67--83, 2021.

\bibitem{Politis_TASLP2020}
A.~Politis, A.~Mesaros, S.~Adavanne, T.~Heittola, and T.~Virtanen,
\newblock ``Overview and evaluation of sound event localization and detection in {DCASE} 2019,''
\newblock {\em IEEE/ACM Transactions on Audio, Speech, and Language Processing {\rm (}TASLP{\rm )}}, vol. 29, pp. 684--–698, Dec. 2020.

\bibitem{Hua_Wei_ESR2018}
Z.~Hua-Wei, H.~Hao, Z.~Zhihui, W.~Yukai, and Y.~Oong,
\newblock ``Reverse time migration: A prospect of seismic imaging methodology,''
\newblock {\em Earth-Science Reviews}, vol. 179, pp. 207--227, 2018.

\bibitem{Kang_sensors2023}
S.~Kang and U.~Jang,
\newblock ``Frequency-domain reverse-time migration with analytic {G}reen’s function for the seismic imaging of shallow water column structures in the arctic ocean,''
\newblock {\em Sensors}, vol. 23, no. 14, pp. 1--15, 2023.

\bibitem{ozdemir_JASA2022}
{\"O}.~{\"O}zg{\"u}r, Y.~Hazel, U.~Y. Ece, E.~Bari{\c{s}}, and E.~Nihal,
\newblock ``{G}reen's functions for a layered high-contrast acoustic media,''
\newblock {\em The Journal of the Acoustical Society of America {\rm (}JASA{\rm )}}, vol. 151, no. 6, pp. 3676--3684, 2022.

\bibitem{gong_interspeech_2021}
Y.~Gong, Y.~Chung, and J.~Glass,
\newblock ``{AST: Audio Spectrogram Transformer},''
\newblock {\em Proc. INTERSPEECH}, pp. 571--575, 2021.

\bibitem{piczak_ACMM2015}
J.~Piczak,
\newblock ``{ESC}: Dataset for environmental sound classification,''
\newblock {\em Proc. the 23rd {Annual ACM Conference} on {Multimedia} {\rm (}ACMM{\rm )}}, pp. 1015--1018, 2015.

\bibitem{koutini_interpseech2022}
K.~Koutini, J.~Schl{\"{u}}ter, H.~Eghbal{-}zadeh, and G.~Widmer,
\newblock ``Efficient training of audio transformers with patchout,''
\newblock {\em Proc. INTERSPEECH}, pp. 2753--2757, 2022.

\bibitem{Elizalde_icassp2022}
B.~Elizalde, S.~Deshmukh, M.~Al Ismail, and H.~Wang,
\newblock ``{CLAP}: learning audio concepts from natural language supervision,''
\newblock {\em Proc. {IEEE} International Conference on Acoustics, Speech and Signal Processing {\rm (}ICASSP{\rm )}}, pp. 1--5, 2023.

\bibitem{Chen_icml2023}
S.~Chen, Y.~Wu, C.~Wang, S.~Liu, D.~Tompkins, Z.~Chen, and F.~Wei,
\newblock ``{BEAT}s: Audio pre-training with acoustic tokenizers,''
\newblock {\em Proc. International conference on machine learning {\rm (}ICML{\rm )}}, pp. 5178--5193, 2023.

\bibitem{imoto_EUSIPCO2021}
K.~Imoto,
\newblock ``Acoustic scene classification using multichannel observation with partially missing channels,''
\newblock {\em Proc. European Signal Processing Conference {\rm (}EUSIPCO{\rm )}}, pp. 875--879, 2021.

\bibitem{Dosovitskiy_iclr2021}
A.~Dosovitskiy, L.~Beyer, A.~Kolesnikov, D.~Weissenborn, X.~Zhai, T.~Unterthiner, M.~Dehghani, M.~Minderer, G.~Heigold, S.~Gelly, J.~Uszkoreit, and N.~Houlsby,
\newblock ``An image is worth 16x16 words: Transformers for image recognition at scale,''
\newblock {\em Proc. International Conference on Learning Representations {\rm (}ICLR{\rm )}}, pp. 1--21, 2021.

\bibitem{Kingma_ICLR2015}
D.~Kingma and J.~Ba,
\newblock ``Adam: A method for stochastic optimization,''
\newblock {\em Proc. International Conference on Learning Representations {\rm (}ICLR{\rm )}}, 2015.

\end{thebibliography}
\end{document}